\documentclass[%
 reprint,
superscriptaddress,
 amsmath,amssymb,
 aps,
]{revtex4-2}

\usepackage{graphicx}% Include figure files
\usepackage{dcolumn}% Align table columns on decimal point
\usepackage{bm}% bold math
\usepackage{mathrsfs}
\usepackage{float}
\usepackage{hyperref}% add hypertext capabilities
\usepackage{xcolor}
\begin{document}

\preprint{APS/123-QED}

\title{A Raman heterodyne determination of the magnetic anisotropy for the ground and optically excited states of Y$_{2}$SiO$_{5}$ doped with Sm$^{3+}$}

\author{N. L. Jobbitt}
\author{J.-P. R. Wells}
\email{Corresponding author: jon-paul.wells@canterbury.ac.nz}
\author{M. F. Reid}
\affiliation{The School of Physical and Chemical Sciences, University of Canterbury, PB4800 Christchurch 8140, New Zealand}
\affiliation{The Dodd-Walls Centre for Photonic and Quantum Technologies, New Zealand}

\author{J. J. Longdell}
\affiliation{The Department of Physics, University of Otago, PB56 Dunedin 9054, New Zealand}
\affiliation{The Dodd-Walls Centre for Photonic and Quantum Technologies, New Zealand}

\date{\today}% It is always \today, today,

\begin{abstract}

We present the full magnetic g tensors of the $^{6}$H$_{5/2}$Z$_{1}$ and $^{4}$G$_{5/2}$A$_{1}$ electronic states for both crystallographic sites in Sm$^{3+}$:Y$_{2}$SiO$_{5}$, deduced through the use of Raman heterodyne spectroscopy performed along 9 different crystallographic directions. The maximum principle g values were determined to be 0.447 (site 1) and 0.523 (site 2) for the ground state and 2.490 (site 1) and 3.319 (site 2) for the excited state. The determination of these g tensors provide essential spin Hamiltonian parameters that can be utilized in future magnetic and hyperfine studies of Sm$^{3+}$:Y$_{2}$SiO$_{5}$, with applications in quantum information storage and communication devices.

\end{abstract}

\maketitle

%\tableofcontents

\section{\label{sec:intro}Introduction}

Lanthanide-doped insulating crystals serve as appealing candidates for the realization of quantum information storage and communication devices. Recently, demonstrations of optical quantum memories and quantum gate implementations have been achieved \cite{Zhong1392, cit:de_riedmatten, Ran_i__2017, PhysRevA.69.032307, PhysRevA.77.022307}. Y$_{2}$SiO$_{5}$ is the host of choice in the realization of these devices owing to the low nuclear spins of the constituent ions, with $^{89}$Y being the only isotope with a non-zero nuclear spin (I = 1/2) that naturally occurs with considerable abundance \cite{PhysRevA.68.012320,FRAVAL2004347}. This reduces the spin-flips induced by the neighbouring ions leading to the long coherence times exhibited by such materials, with observations of coherence times exceeding 1 minute for Pr$^{3+}$:Y$_{2}$SiO$_{5}$ and 6 hours for Eu$^{3+}$:Y$_{2}$SiO$_{5}$ \cite{PhysRevLett.111.033601,Zhong}. These coherence times were obtained through the use of the Zero-First-Order-Zeeman (ZEFOZ) technique which utilizes an external magnetic field to minimize dephasing induced by spin flips on neighbouring host lattice ions. The field points at which this occurs are known as ZEFOZ points, which are avoided crossings of the hyperfine levels that exist within the Zeeman-hyperfine structure of lanthanide-doped materials. ZEFOZ points are difficult to find experimentally but can be computationally predicted, for example, through the use of the spin Hamiltonian \cite{PhysRevB.66.035101}. The studies that resulted in the observed coherence times were enabled by previous studies that determined spin Hamiltonian parameters for Pr$^{3+}$:Y$_{2}$SiO$_{5}$ and Eu$^{3+}$:Y$_{2}$SiO$_{5}$ \cite{PhysRevLett.111.033601,Zhong,PhysRevB.66.035101, PhysRevB.85.014429,cit:Longdell2}.

Kramers systems provide an appealing alternative to non-Kramers ions in applications of quantum information storage and communication devices owing to ions such as Sm$^{3+}$ and Er$^{3+}$ having large hyperfine splittings relative to Pr$^{3+}$ and Eu$^{3+}$ \cite{horvath, PhysRevLett.111.033601,Zhong}. This allows for larger memory bandwidths within these hyperfine transitions whilst still obtaining reasonably long coherence times. Previously, a hyperfine coherence time of 1.3 s was obtained for Er$^{3+}$:Y$_{2}$SiO$_{5}$ through the use of high magnetic fields strengths \cite{Ran_i__2017}.  Furthermore, studies have obtained hyperfine coherence times of 1 ms and 1.48 ms for Yb$^{3+}$:Y$_{2}$SiO$_{5}$ and Er$^{3+}$:Y$_{2}$SiO$_{5}$ respectively without the need of applying an external magnetic field \cite{cit:ortu, cit:Rakonjac}. These studies were enabled by previously determined spin Hamiltonian parameters \cite{PhysRevB.94.155116, PhysRevB.74.214409}. In particular, Sm$^{3+}$:Y$_{2}$SiO$_{5}$ provides a not yet investigated alternative in such applications thanks to the small ground state g values, which results in a relative insensitivity to magnetic field fluctuations. Additionally, Sm$^{3+}$ has a multitude of isotopes with 71.2 \% of all naturally occurring Sm$^{3+}$ possessing zero nuclear spin, and two isotopes, $^{147}$Sm and $^{149}$Sm, with natural abundances of 15.0 \% and 13.8 \% respectively, both have a nuclear spin of 7/2. This gives rise to the possibility of multiple ZEFOZ points within the Sm$^{3+}$:Y$_{2}$SiO$_{5}$ system for both of the non-zero nuclear spin isotopes \cite{Jobbitt2019}.

We report on the determination of the magnetic g tensors of the $^{6}$H$_{5/2}$Z$_{1}$ and $^{4}$G$_{5/2}$A$_{1}$ states for both sites of Sm$^{3+}$:Y$_{2}$SiO$_{5}$ through the use of Raman heterodyne spectroscopy, focusing on the nuclear spin zero isotopes of Sm$^{3+}$. The $^{4}$G$_{5/2}$A$_{1}$ state, located at $\sim$560 nm, is of particular interest as it is the only emitting state in Sm$^{3+}$:Y$_{2}$SiO$_{5}$, owing to non-radiative relaxation between all other states, and is readily accessible with conventional visible lasers \cite{cit:jobbitt2}. Raman heterodyne spectroscopy is a widely used spectroscopic technique used in probing the hyperfine structure of lanthanide-doped systems, as first demonstrated in the detection of nuclear magnetic resonance of Pr$^{3+}$:LaF$_{3}$ \cite{cit:mlynek, cit:wong}. Recent studies have shown the ability to characterize many non-Kramers in addition to Kramers systems through the determination of spin Hamiltonian parameters \cite{cit:Longdell, cit:Longdell2, cit:Fernandez}. The ability to determine such spin Hamiltonian parameters, including the g tensors determined in this study, are essential precursors in the development of quantum information storage and communication devices.

\section{\label{sec:experimental}Experimental}

Y$_{2}$SiO$_{5}$ is a monoclinic silicate crystal having space group C$^{6}_{2h}$ \cite{cit:maksimov}. The lattice constants of Y$_{2}$SiO$_{5}$ are a = 10.4103 \AA, b = 6.7212 \AA, c = 12.4905 \AA, and $ \beta $  = 102$^{\circ}$39'. Here the crystallographic b axis corresponds to the C$_{2}$ rotation axis and the crystallographic a and c axes are located in the mirror plane which is perpendicular to the crystallographic b axis. Each unit cell of Y$_{2}$SiO$_{5}$ is composed of eight Y$_{2}$SiO$_{5}$ molecules, with each molecule containing two substitutional Y\textsuperscript{3+} sites, denoted site 1 and site 2. Both of these sites have C\textsubscript{1} symmetry and are distinguished by their coordination numbers of six and seven respectively. Following the convention of Li et al. we define the optical extinction axes as D$_{1}$ and D$_{2}$ which are located in the a-b mirror plane and are perpendicular to each other in addition to the crystallographic b axis \cite{cit:li}. Additionally, each site of Y$_{2}$SiO$_{5}$ also contains two sub-sites which are related by a 180$^{\circ}$ rotational symmetry and respond differently when a magnetic field is applied outside of the D$_{1}$-D$_{2}$ plane or the b axis \cite{cit:Sun}.

The sample used in this study was grown in the X$_{2}$ phase of Y$_{2}$SiO$_{5}$ using the Czochralski process by Scientific Materials Inc. (Bozeman, USA), with a Sm$^{3+}$ dopant concentration of 0.5 molar \%. The crystal had dimensions of (5.1 $\pm$ 0.1) mm along the D$_{1}$ axis, (4.9 $\pm$ 0.1) mm along the D$_{2}$ axis, and (6.0 $\pm$ 0.2) mm along the crystallographic b axis.

In order to perform Raman heterodyne spectroscopy, the sample was attached to an aluminium sample holder that also included a 4 loop copper coil. This coil allows RF coupling to the sample up to approximately 70 MHZ.

\begin{figure}[H]
	\centering
	\includegraphics[width=0.48\textwidth]{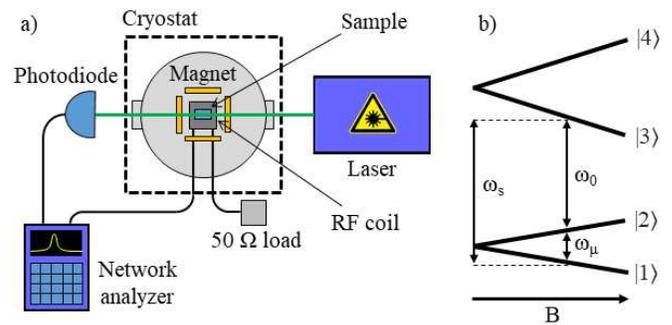}
	\caption{a) Experimental setup for Raman Heterodyne spectroscopy. b) Energy level diagram relevant to Raman Heterodyne spectroscopy. The Zeeman transition of the ground and excited states are driven by a RF field, $\omega_{\mu}$, whereas the transition between the ground and excited states is driven by an optical frequency field, $\omega_{0}$. The resulting optical field has a frequency, $\omega_{s}$, which is equal to the sum of the incident optical frequency field and RF frequencies.}
	\label{fig:experimental_setup}
\end{figure}

Raman heterodyne spectroscopy involves coupling a RF excitation source to an optical source supplied by a single frequency laser. The experimental setup is given in Figure \ref{fig:experimental_setup} a). Upon the application of a magnetic field, the Kramers degeneracy is removed allowing the ground and excited states to each split into two Zeeman states. Figure \ref{fig:experimental_setup} b) depict the energy level diagram relevant to Raman Heterodyne spectroscopy.

The sample holder was screwed into a cryostat which allowed the sample to be cooled to 3.2 K. Within the cryostat, located around the sample holder is a HTS-110 Ltd. liquid nitrogen cooled superconducting vector magnet.

The RF field was provided by a network analyzer. When the RF field is resonant to a Zeeman transistion, the resulting optical field leaving the sample is composed of the two optical frequencies, $\omega_{s}$ and $\omega_{0}$, with a beat frequency of $\omega_{\mu}$. This signal was detected by a photodiode and was then measured by the network analyzer. The network analyzer sweeps through a RF field range while simultaneously measuring the beat signal of the resulting optical field.

Zeeman absorption spectroscopy was performed using a Bruker Vertex 80 Fourier transform infrared (FTIR) spectrometer having a maximum apodized resolution of 0.075 cm$^{-1}$. The sample was thermally attached to a copper mount which was then screwed into the bore of a 4 T Oxford Instruments superconducting solenoid built into a home-built helium cryostat.

\begin{figure}[H]
	\centering
	\includegraphics[width=0.3\textwidth]{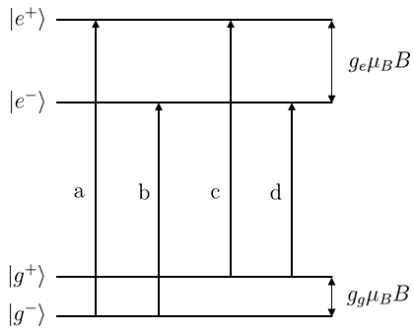}
	\caption{The four transitions seen for each spectral line when an external magnetic is applied along the b-axis or the D$_{1}$-D$_{2}$ plane, lifting Kramers degeneracy.}
	\label{fig:splittings}
\end{figure}

When a magnetic field is applied to Sm$^{3+}$:Y$_{2}$SiO$_{5}$, the Kramers degeneracy is lifted resulting in the splitting of each state into two. This results in each electronic transition to split into four transitions when the magnetic field is applied along the b-axis or the D$_{1}$-D$_{2}$ plane as depicted in Figure \ref{fig:splittings}. In a general magnetic field direction, due to the two magnetic inequivalent orientations of Y$_{2}$SiO$_{5}$, each transition splits into eight transitions instead. 

From Figure \ref{fig:splittings} we see that the ground state g values can be expressed as:

\begin{equation}
g_{g} = \frac{E_{a} - E_{c}}{\mu_{B}B} = \frac{E_{b} - E_{d}}{\mu_{B}B}
\label{eq:g_ground}
\end{equation}

And the excited state g values can be expressed as:

\begin{equation}
g_{e} = \frac{E_{a} - E_{b}}{\mu_{B}B} = \frac{E_{c} - E_{d}}{\mu_{B}B}
\label{eq:g_excited}
\end{equation}

In both of the above equations, $B$ is the applied magnetic field strength and $\mu_{B}$ is the Bohr Magneton.

\section{\label{sec:results}Results and Discussion}

\begin{figure}[h!]
	\centering
	\includegraphics[width=0.48\textwidth]{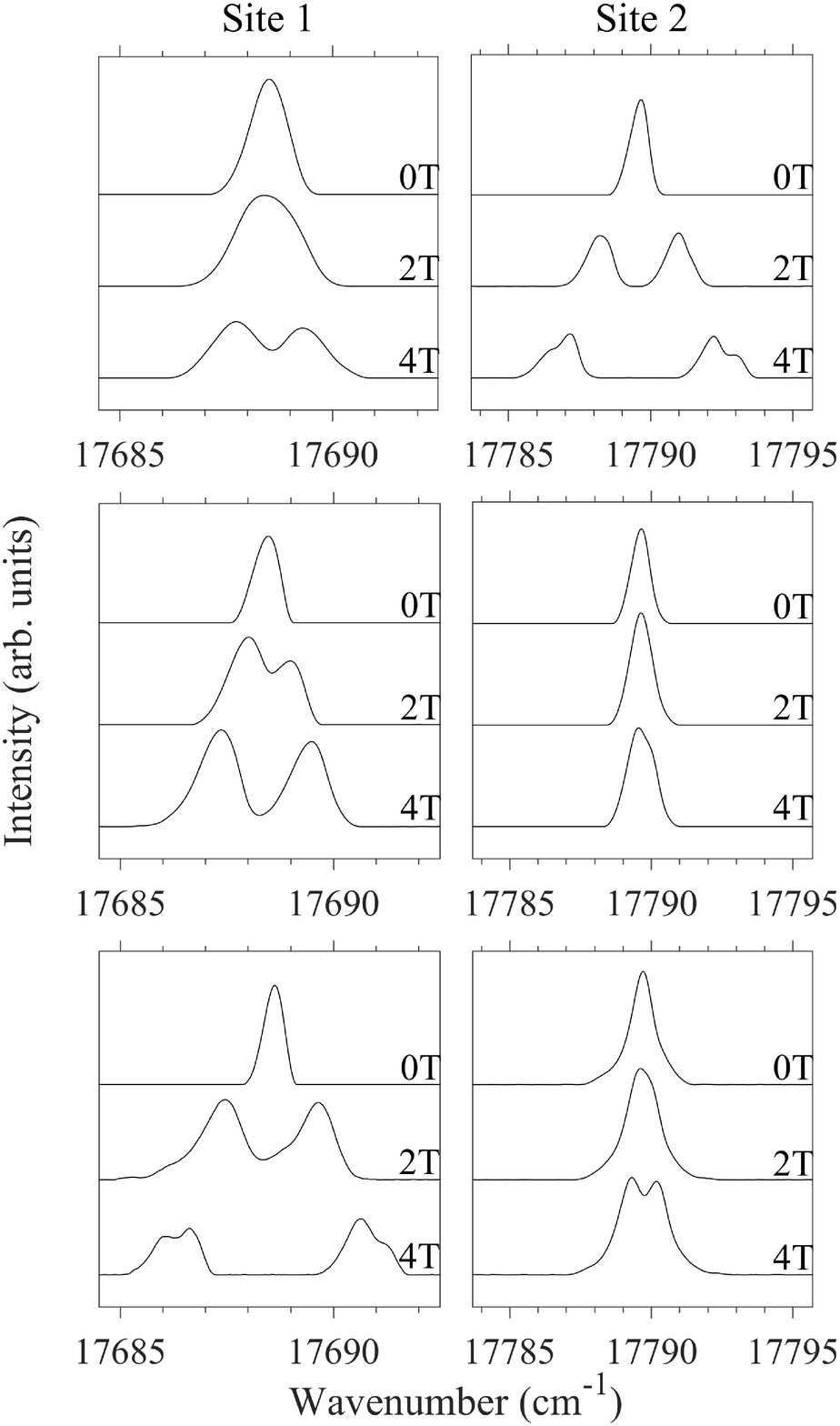}
	\caption{4.2 K Zeeman spectra of the $^{6}$H$_{5/2}$Z$_{1} \longrightarrow \ ^{4}$G$_{5/2}$A$_{1}$ transition for site 1 (left panels) and site 2 (right panels) in Sm$^{3+}$:Y$_{2}$SiO$_{5}$ with the magnetic field applied along the D$_{1}$ (top), D$_{2}$ (middle) and b (bottom) axes. Spectra obtained at 0 T, 2 T and 4 T are displayed.}
	\label{fig:zeeman}
\end{figure}

Initially Zeeman absorption spectroscopy was performed on Sm$^{3+}$:Y$_{2}$SiO$_{5}$ which, owing to the geometry of our setup, limited the ability of determining the g values to only along the D$_{1}$, D$_{2}$ and b axes of Y$_{2}$SiO$_{5}$. Figure \ref{fig:zeeman} shows the optical Zeeman absorption spectra for the $^{6}$H$_{5/2}$Z$_{1} \longrightarrow \ ^{4}$G$_{5/2}$A$_{1}$ transitions with magnetic fields applied along the extinction axes of the Sm$^{3+}$:Y$_{2}$SiO$_{5}$ sample. The small ground state splitting is unresolvable in some directions. The g values that could be measured using optical absorption were calculated using equations (\ref{eq:g_ground}) and (\ref{eq:g_excited}) and are summarized in Table \ref{tab:zeeman_gvalues}. The g values were assigned as belonging to either the ground or excited state through comparison to other lines found in absorption, (not included here for brevity), as every transition shares a common ground state.

\begin{table}[h!]
	\centering
	
	\caption{g values obtained through Zeeman spectroscopy along the extinction axes of Y$_{2}$SiO$_{5}$ as shown in Figure \ref{fig:zeeman}. Values marked with a '--' could not be identified due to the ground state splittings being too small to resolve. Each g value has an uncertainty of 0.1, this reflects the uncertainty in the peak position of the spectral lines.}
	\begin{ruledtabular}
	\begin{tabular}{ cccccc }
	& \multicolumn{2}{c}{Site 1} & & \multicolumn{2}{c}{Site 2} \\\cline{2-3}\cline{5-6}
	Direction        & Ground       & Excited      & & Ground       & Excited\\\hline
	D$_{1}$          & --           & 1.0          & & 0.4          & 3.2 \\
	D$_{2}$          & --           & 1.2          & & --           & 0.3 \\
	b                & 0.3          & 2.4          & & --           & 0.6 \\
	\end{tabular}
	\end{ruledtabular}
	\label{tab:zeeman_gvalues}
\end{table}

Zeeman spectroscopy proved to be insufficient to derive the full g tensor of any one state as the full g tensor has 6 independent components and therefore requires g values to be obtained along at least 6 different directions. Raman heterodyne spectroscopy was performed at low magnetic field strengths along 9 different directions in order to determine the g tensors for the ground $^{6}$H$_{5/2}$Z$_{1}$ and optically excited $^{4}$G$_{5/2}$A$_{1}$ states of both sites. All spectra were obtained at 3.5 K and the magnetic field was applied down the three extinction axes or at 45$^{\circ}$ between two of the axes resulting in 9 different directions.

\begin{figure}[H]
	\centering
	\includegraphics[width=0.48\textwidth]{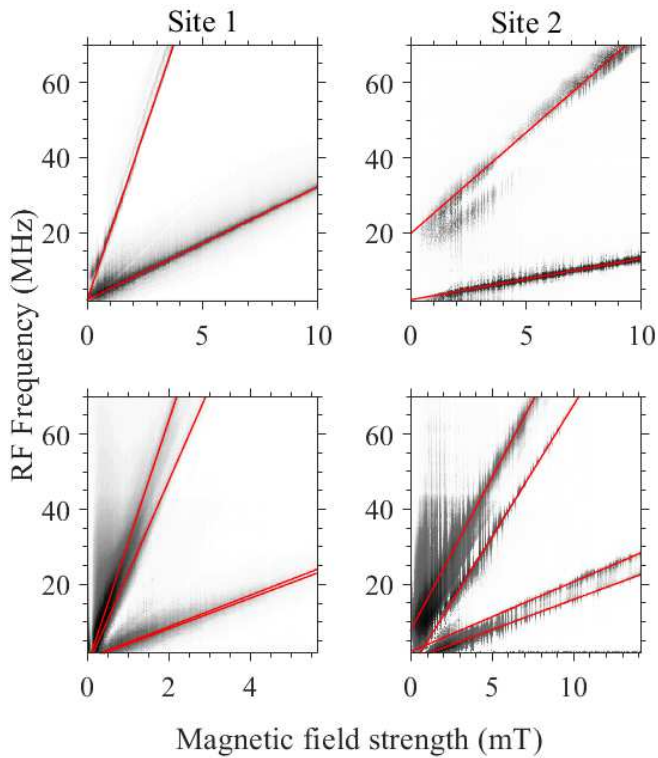}
	\caption{3.5 K Raman heterodyne signal for the site 1 (left) and site 2 (right), $^{6}$H$_{5/2}$Z$_{1} \longrightarrow \ ^{4}$G$_{5/2}$A$_{1}$ transition of Sm$^{3+}$:Y$_{2}$SiO$_{5}$ along the D$_{2}$ axis (top) and D$_{2}$,b direction (bottom). The red lines represents the linear least-squares fits to the experimental data. The laser excitation wavelength was 17 688.6 cm$^{-1}$ for site 1 and 17 789.6 cm$^{-1}$ for site 2. The zero offsets are due to residual magnetic fields that are always present in the magnet.}
	\label{fig:RH}
\end{figure}

Figure \ref{fig:RH} shows representative Raman heterodyne spectra for the $^{6}$H$_{5/2}$Z$_{1} \longrightarrow \ ^{4}$G$_{5/2}$A$_{1}$ transition for site 1 (left panels) and site 2 (right panels) in Sm$^{3+}$:Y$_{2}$SiO$_{5}$ along the D$_{2}$ direction (top panels) and the D$_{2}$,b direction (bottom panels), obtained at 3.5 K. Note that the magnetic field strength varies between spectra and that the zero offsets are due to a stray magnetic field. The Zeeman transitions of interest are the linear lines visible in each spectrum. The top spectra has a magnetic field applied along the D$_{2}$ axis and show two transitions, one representing the splitting of the ground state and the other representing the splitting of the excited state. The presence of four transitions in the D$_{1}$,$\pm$b and D$_{2}$,$\pm$b directions as represented in the bottom two panels in Figure \ref{fig:RH} are due to the two magnetically inequivalent subsites of Y$_{2}$SiO$_{5}$. The additional structure seen in Figure \ref{fig:RH} we attribute to interactions that arise from the relatively high concentration of our sample (0.5 molar \%) in addition to unresolved hyperfine structure from both of the non-zero nuclear spin isotopes.

The g values of the $^{6}$H$_{5/2}$Z$_{1}$ ground and $^{4}$G$_{5/2}$A$_{1}$ excited states were determined and are summarized in Table \ref{tab:RH_gvalues}. These g values agree with those found in Zeeman absorption spectroscopy as given in Table \ref{tab:zeeman_gvalues}. Each state in the D$_{1}$,$\pm$ b and D$_{2}$,$\pm$ b directions have two related g values which arises from the two magnetically inequivalent subsites. In the case of Y$_{2}$SiO$_{5}$, the D$_{1}$,+b and D$_{1}$,-b directions in addition to the the D$_{2}$,+b and the D$_{2}$,-b directions are degenerate, therefore the average of the two g values were used in determining the g tensors. Preliminary crystal-field analyses performed on Sm$^{3+}$:Y$_{2}$SiO$_{5}$ have shown that the g values for the ground state are significantly smaller than those of the excited state in most directions, and therefore distinguishing their g values is trivial \cite{Jobbitt2019}. The exception to this are the g values in the D$_{2}$, b and D$_{2}$,$\pm$b directions for site 2 where both the ground and excited states have g values significantly less than one. These g values were classified as belonging to that of either the ground or excited state by constructing a g tensor for every remaining combination of g values and determining the g tensor that provided the closest agreement to the experimental data.

\begin{table*}
	\centering
	
	\caption{Experimentally determined g values of the $^{6}$H$_{5/2}$Z$_{1}$ ground and $^{4}$G$_{5/2}$A$_{1}$ excited states, represented in Figure \ref{fig:RH}, along each magnetic field direction for both sites of Sm$^{3+}$:Y$_{2}$SiO$_{5}$. The numbers in parentheses are the uncertainties of their respective g values.} 
	\begin{ruledtabular}
	\begin{tabular}{ cccccc }
	& \multicolumn{2}{c}{Site 1} & & \multicolumn{2}{c}{Site 2} \\\cline{2-3}\cline{5-6}
	Direction        & Ground       & Excited      & & Ground       & Excited\\\hline
	D$_{1}$          & 0.36 (0.04) & 1.07  (0.31)   & & 0.52 (0.05)        & 3.29 (0.51)\\
	D$_{2}$          & 0.21 (0.09)        & 1.29   (0.21)  & & 0.078 (0.005)        & 0.38 (0.11)\\
	b                & 0.39 (0.02) & 2.46 (0.19)   & & 0.15 (0.01)        & 0.77 (0.11) \\
	D$_{1}$,+D$_{2}$ & 0.31 (0.05) & 0.94 (0.14) & & 0.34  (0.05) & 2.13 (0.16) \\
	D$_{1}$,-D$_{2}$ & 0.39 (0.18) & 1.43 (0.34)  & & 0.41  (0.05) & 2.61 (0.09) \\
	D$_{1}$,$\pm$ b  & 0.28, 0.50 (0.01, 0.15) & 1.77, 2.16 (0.08, 0.42) & & 0.27, 0.44 (0.05, 0.02) & 2.40, 2.42 (0.24, 0.35) \\
	D$_{2}$,$\pm$ b  & 0.29, 0.29 (0.12, 0.12) & 1.68, 2.29 (0.23, 0.49) & & 0.10, 0.146 (0.01, 0.005) & 0.50, 0.66 (0.08, 0.06) \\

	\end{tabular}
	\end{ruledtabular}
	\label{tab:RH_gvalues}
\end{table*}

Following the conventions of Weil et. al. we relate the g values, $g$, with a magnetic field applied along an arbitrary direction $\mathbf{n}$ to the g tensor, $\mathbf{g}$, through the following relationship \cite{cit:weil}:

\begin{equation}
       g^{2}(\mathbf{n})  = \mathbf{n}^{T} \cdot (\mathbf{g}\cdot \mathbf{g}^{T}) \cdot \mathbf{n}
    \label{eq:gtensor}
\end{equation}

For a particular magnetic field direction, Equation (\ref{eq:gtensor}) is transformed to:

\begin{equation}
       g^{2}(\mathbf{n}_{\alpha})  = (\mathbf{g}\cdot \mathbf{g}^{T})_{\alpha\alpha}
    \label{eq:transformed_gtensor}
\end{equation}

and

\begin{equation}
       g^{2}(\mathbf{n}_{\alpha\pm\beta})  = \frac{1}{2} \Big[ (\mathbf{g}\cdot \mathbf{g}^{T})_{\alpha\alpha} + (\mathbf{g}\cdot \mathbf{g}^{T})_{\beta\beta} \pm 2(\mathbf{g}\cdot \mathbf{g}^{T})_{\alpha\beta} \Big]
    \label{eq:transformed_gtensor_2}
\end{equation}

Here $\mathbf{n}_{\alpha}$ and $\mathbf{n}_{\beta}$ are the basis vectors of the Cartesian coordinate system and $\mathbf{n}_{\alpha\pm\beta}$ are the unit vectors in the $\mathbf{n}_{\alpha} \pm \mathbf{n}_{\beta}$ directions. From Equations (\ref{eq:transformed_gtensor}) and (\ref{eq:transformed_gtensor_2}) the off-diagonal components of the g tensor can be expressed as:

\begin{equation}
    (\mathbf{g}\cdot \mathbf{g}^{T})_{\alpha\beta} = \frac{g^{2}(\mathbf{n}_{\alpha+\beta}) - g^{2}(\mathbf{n}_{\alpha-\beta})}{2}
    \label{eq:offdiag_gtensor}
\end{equation}

Using Equations (\ref{eq:transformed_gtensor}) and (\ref{eq:offdiag_gtensor}) the full g tensors can be determined. The g tensor is symmetric and therefore has 6 independent components that are required to be determined. Equations (\ref{eq:gtensors_site1_ground} -- \ref{eq:gtensors_site2_excited}) show the g tensors of the $^{6}$H$_{5/2}$Z$_{1}$ ground ($\mathbf{g}_{g1}$ for site 1 and $\mathbf{g}_{g2}$ for site 2) and $^{4}$G$_{5/2}$A$_{1}$ excited ($\mathbf{g}_{e1}$ for site 1 and $\mathbf{g}_{e2}$ for site 2) states in Sm$^{3+}$:Y$_{2}$SiO$_{5}$ that provides the closest agreement to the experimental data.

\begin{equation}
       \mathbf{g}_{g1}  = \begin{pmatrix}
        0.351 & -0.016 &  0.078\\
        -0.016 & 0.209 & -0.007\\
        0.078 & -0.007 & 0.382
        \end{pmatrix}
    \label{eq:gtensors_site1_ground}
\end{equation}

\begin{equation}
        \mathbf{g}_{e1}  = \begin{pmatrix}
        1.025 & -0.257 & 0.166\\
        -0.257 & 1.248 & 0.202\\
        0.166 & 0.202 & 2.446
        \end{pmatrix}
    \label{eq:gtensors_site1_excited}
\end{equation}

\begin{equation}
       \mathbf{g}_{g2}  = \begin{pmatrix}
        0.512 & -0.040 & -0.054\\
        -0.040 & 0.067 & 0.009\\
        -0.054 & 0.009 & 0.135
        \end{pmatrix}
    \label{eq:gtensors_site2_ground}
\end{equation}

\begin{equation}
        \mathbf{g}_{e2}  = \begin{pmatrix}
        3.264 & -0.311 & 0.262\\
        -0.311 & 0.183 & 0.125\\
        0.262 & 0.125 & 0.714
        \end{pmatrix}
    \label{eq:gtensors_site2_excited}
\end{equation}

\begin{figure}[H]
	\centering
	\includegraphics[width=0.4\textwidth]{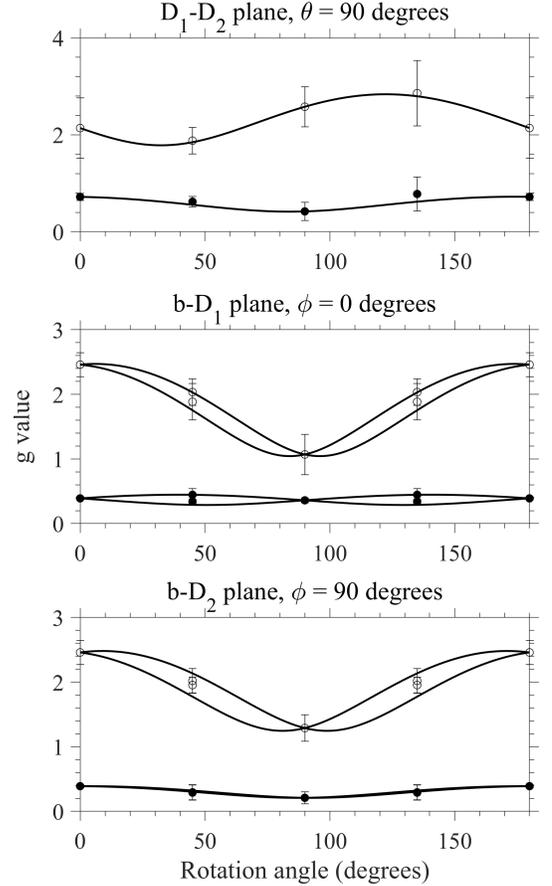}
	\caption{g value rotation curves in the a) D$_{1}$-D$_{2}$ ($\theta = 90^{\circ}$) plane, b) b-D$_{1}$ ($\phi = 0^{\circ}$) plane and c) b-D$_{2}$ ($\phi = 90^{\circ}$) plane for the $^{6}$H$_{5/2}$Z$_{1}$ ground state (solid circles) and the G$_{5/2}$A$_{1}$ excited state (hollow circles) for site 1 in Sm$^{3+}$:Y$_{2}$SiO$_{5}$. The solid lines are simulated g values based on the determined g tensors given in Equations (\ref{eq:gtensors_site1_ground}) and (\ref{eq:gtensors_site1_excited}).}
	\label{fig:rollercoaster_site1}
\end{figure}

\begin{figure}[H]
	\centering
	\includegraphics[width=0.4\textwidth]{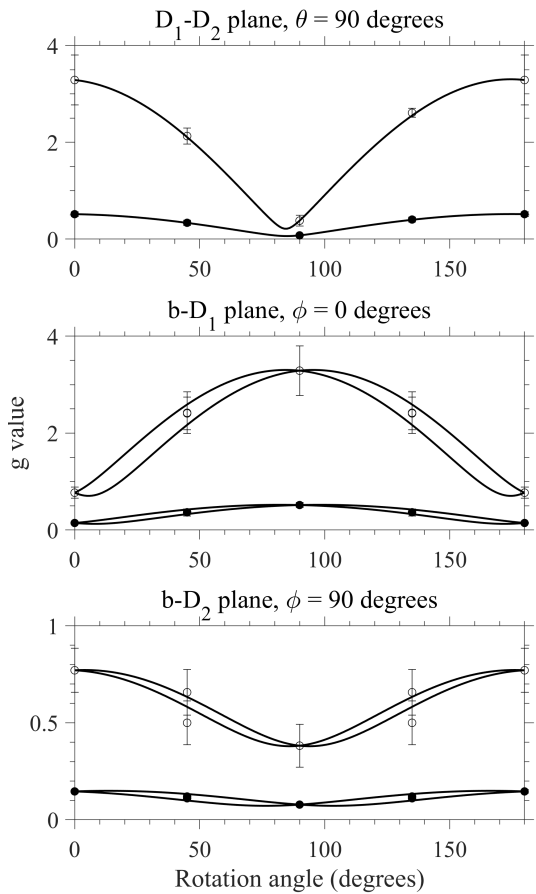}
	\caption{g value rotation curves in the a) D$_{1}$-D$_{2}$ ($\theta = 90^{\circ}$) plane, b) b-D$_{1}$ ($\phi = 0^{\circ}$) plane and c) b-D$_{2}$ ($\phi = 90^{\circ}$) plane for the $^{6}$H$_{5/2}$Z$_{1}$ ground state (solid circles) and the G$_{5/2}$A$_{1}$ excited state (hollow circles) for site 2 in Sm$^{3+}$:Y$_{2}$SiO$_{5}$. The solid lines are simulated g values based on the determined g tensors given in Equations (\ref{eq:gtensors_site2_ground}) and (\ref{eq:gtensors_site2_excited}).}
	\label{fig:rollercoaster_site2}
\end{figure}

Figures \ref{fig:rollercoaster_site1} and  \ref{fig:rollercoaster_site2} depict the angular dependence of the g values for the $^{6}$H$_{5/2}$Z$_{1}$ ground and $^{4}$G$_{5/2}$A$_{1}$ excited states of sites 1 and 2 respectively. The top panel depict a rotation in the D$_{1}$-D$_{2}$ ($\theta = 90^{\circ}$) plane with $\phi = 0^{\circ}$ corresponding to the D$_{1}$ axis and $\phi = 90^{\circ}$ corresponding to the D$_{2}$ axis. The middle panel depict a rotation in the b-D$_{1}$ ($\phi = 0^{\circ}$) plane with $\theta = 0^{\circ}$ corresponding to the b axis and $\theta = 90^{\circ}$ corresponding to the D$_{1}$ axis. The bottom panel depict a rotation in the b-D$_{2}$ ($\phi = 90^{\circ}$) plane with $\theta = 0^{\circ}$ corresponding to the b axis and $\theta = 90^{\circ}$ corresponding to the D$_{2}$ axis. The experimental data is depicted as solid circles for the ground state and as hollow circles for the excited state.

Table \ref{tab:principleg} depicts the principal axes and related direction cosines obtained by diagonalising the g tensors given in Equations (\ref{eq:gtensors_site1_ground}) --- (\ref{eq:gtensors_site2_excited}). The principal axes are the eigenvectors of the g tensor while the principal g values are their corresponding eigenvalues. The principal axes are labelled $x'$, $y'$ and $z'$ with the maximum and minimum g value directions labelled as $z'$ and $x'$ respectively.

\begin{table*}
	\caption{Principal g values and direction cosines of the ground and excited state g tensors for both sites in Sm$^{3+}$:Y$_{2}$SiO$_{5}$. The principal axes are labelled $x'$, $y'$ and $z'$ with the maximum and minimum g values directions labelled as $z'$ and $x'$ respectively.}
    \begin{ruledtabular}
	\begin{tabular}{ ccccccccccc }
	& & \multicolumn{4}{c}{Ground state} & & \multicolumn{4}{c}{Excited state} \\\cline{3-6}\cline{8-11}
	
        	&     & Principal g & l      & m      & n      & & Principal g & l     & m      & n       \\\hline
	Site 1  & g$_{z'}$ & 0.447       & 0.635 & -0.066 & 0.770 & & 2.490       & 0.087 & 0.143 & 0.986 \\
	        & g$_{y'}$ & 0.288       & -0.764 &  0.093  & 0.639 & & 1.411       & 0.573 & -0.817 & 0.068 \\
	        & g$_{x'}$ & 0.208       & 0.114 & 0.994 & -0.009 & & 0.818       & 0.815 & 0.559 & -0.152 \\
	        &&&&&&&&&&\\
	Site 2  & g$_{z'}$ & 0.523       & -0.986 & 0.088  & 0.140  & & 3.319       & -0.991 & 0.095 & -0.095 \\
	        & g$_{y'}$ & 0.128       & -0.144 & -0.045 & -0.989 & & 0.730       & -0.068 & 0.259 & 0.964 \\
	        & g$_{x'}$ & 0.063       & 0.081  & 0.995  & -0.057 & & 0.113       & 0.116 & 0.961 & -0.251 \\
				
	\end{tabular}
	\end{ruledtabular}
	\label{tab:principleg}
\end{table*}

\begin{figure*}
	\centering
	\includegraphics[width=0.8\textwidth]{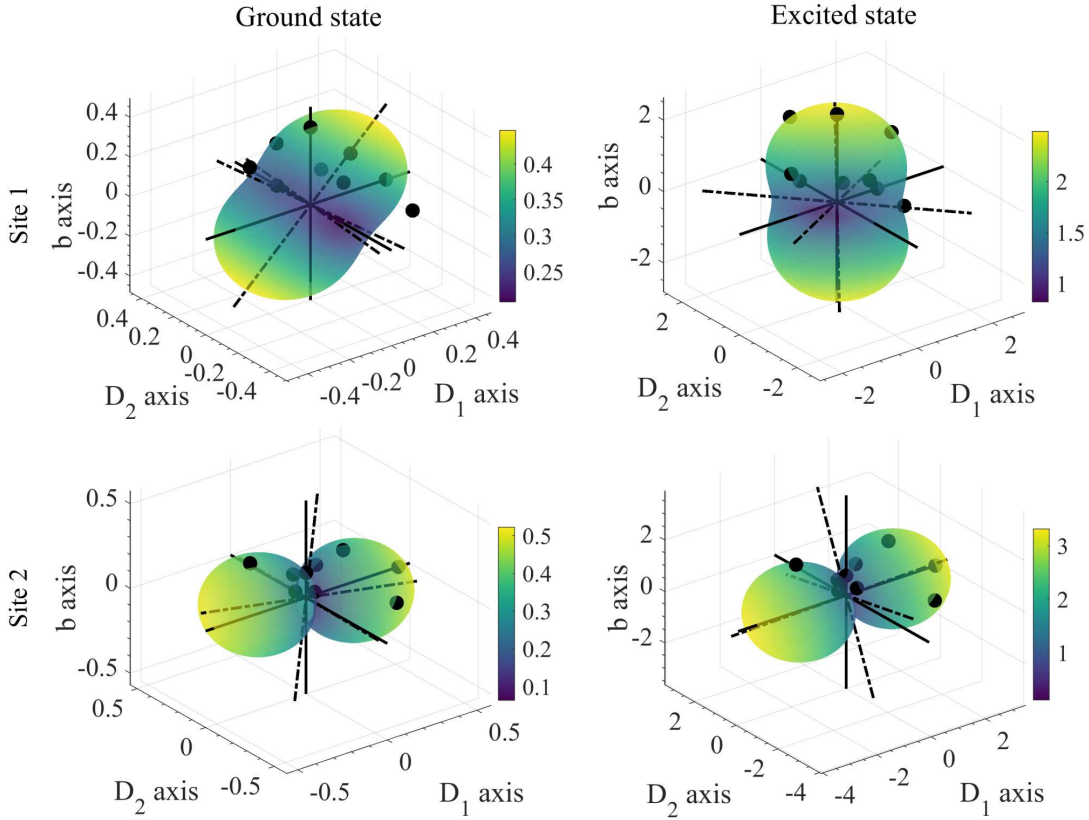}
	\caption{The effective g values of Y$_{2}$SiO$_{5}$ for the ground (left panels) and excited (right panels) for site 1 (top panels) and site 2 (bottom panels). The experimental values are given by the black data points whereas the extinction and principal axes are given as the solid and dash-dotted lines respectively.}
	\label{fig:gtensors}
\end{figure*}

Figure \ref{fig:gtensors} gives a visual representation of the g tensors. The g tensors derived in Equations (\ref{eq:gtensors_site1_ground}) --- (\ref{eq:gtensors_site2_excited}) were used to derive the effective g values in terms of the extinction axes for both the $^{6}$H$_{5/2}$Z$_{1}$ ground state and the $^{4}$G$_{5/2}$A$_{1}$ excited state of Y$_{2}$SiO$_{5}$. The extinction and principal axes are also depicted as the solid and dash-dotted lines respectively.

\section{Conclusion}

We have studied the $^{6}$H$_{5/2}$Z$_{1}$ and $^{4}$G$_{5/2}$A$_{1}$ states for both crystallographic sites in Sm$^{3+}$:Y$_{2}$SiO$_{5}$ through the use of Raman heterodyne spectroscopy. The g values determined along 9 different crystallographic directions were used to determine the full g tensors for these states. Sm$^{3+}$:Y$_{2}$SiO$_{5}$ is an attractive alternative in the development of quantum information storage and communication devices relative to traditional systems such as Pr$^{3+}$:Y$_{2}$SiO$_{5}$ and Eu$^{3+}$:Y$_{2}$SiO$_{5}$. This is due to Sm$^{3+}$ possessing a large hyperfine splitting, multiple non-zero nuclear spin isotopes, and a relative insensitivity to magnetic field fluctuations when compared to Er$^{3+}$:Y$_{2}$SiO$_{5}$, which is a result of the very small ground state g values. This allows for the possibility of a multitude of ZEFOZ points to be experimentally determined and utilized for greater bandwidth quantum information storage and communication devices.

\begin{acknowledgments}
NLJ would like to thank the Dodd-Walls Centre for Photonic and Quantum Technologies for the provision of a PhD studentship. The technical assistance of Mr J. Everts and Ms M. Cormack is gratefully acknowledged.

\end{acknowledgments}

%\bibliography{bibliography}% Produces the bibliography via BibTeX.


\end{document}